\documentclass[runningheads]{llncs}
\usepackage{graphicx}
\usepackage{comment}
\usepackage{amsmath,amssymb} 
\usepackage{color}

\usepackage{multicol}
\usepackage{cases}
\usepackage{overpic}
\usepackage{caption}
\usepackage{subcaption}
\usepackage{multirow}
\usepackage{tabularx}
\usepackage{enumitem}
\usepackage{rotating}
\usepackage{textcomp}

\def\bx{{\bf x}}
\def\by{{\bf y}}

\def\0{{\bf 0}}
\def\1{{\bf 1}}

\def\bH{{\bf H}}

\def\etal{\emph{et al. }}
\def\ie{\emph{i.e. }}

\usepackage{flushend}
\usepackage[pagebackref=true,breaklinks=true,letterpaper=true,colorlinks,bookmarks=false]{hyperref}

\begin{document}
\pagestyle{headings}
\mainmatter
\def\ECCVSubNumber{83}  

\title{Deep Cyclic Generative Adversarial Residual Convolutional Networks for Real Image Super-Resolution} 

%
\titlerunning{SRResCycGAN}

\authorrunning{Rao Muhammad Umer \textit{et al.}}

\author{Rao Muhammad Umer, Christian Micheloni}
\institute{
	University of Udine, Italy. \\
	\scriptsize \email{ engr.raoumer943@gmail.com, christian.micheloni@uniud.it }
}

%
%
\maketitle

\begin{abstract}
Recent deep learning based single image super-resolution (SISR) methods mostly train their models in a clean data domain where the low-resolution (LR) and the high-resolution (HR) images come from noise-free settings (same domain) due to the bicubic down-sampling assumption. However, such degradation process is not available in real-world settings. We consider a deep cyclic network structure to maintain the domain consistency between the LR and HR data distributions, which is inspired by the recent success of CycleGAN in the image-to-image translation applications. We propose the Super-Resolution Residual Cyclic Generative Adversarial Network (SRResCycGAN\footnote{Our code and trained models are publicly available at \url{https://github.com/RaoUmer/SRResCycGAN}}) by training with a generative adversarial network (GAN) framework for the LR to HR domain translation in an end-to-end manner. We demonstrate our proposed approach in the quantitative and qualitative experiments that generalize well to the real image super-resolution and it is easy to deploy for the mobile/embedded devices. In addition, our SR results on the AIM 2020 Real Image SR Challenge datasets demonstrate that the proposed SR approach achieves comparable results as the other state-of-art methods. 
\end{abstract}

\section{Introduction}
The goal of the single image super-resolution (SISR) is to recover the high-resolution (HR) image from its low-resolution (LR) counterpart. SISR problem is a fundamental low-level vision and image processing problem with various practical applications in  satellite imaging, medical imaging, astronomy, microscopy imaging, seismology, remote sensing, surveillance, biometric, image compression, etc. Usually, the SISR is described as a linear forward observation model by the following image degradation process:
\begin{equation}
    \by = \bH * \Tilde{\bx} + \eta,
    \label{eq:degradation_model}
\end{equation}
where $\by$ is an observed LR image, $\bH$ is a \emph{down-sampling operator} (usually bicubic) that convolves with an HR image $\Tilde{\bx}$ and resizes it by a scaling factor $s$, and $\eta$ is considered as an additive white Gaussian noise with standard deviation $\sigma$. However, in real-world settings,  $\eta$ also accounts for all possible errors during the image acquisition process that include inherent sensor noise, stochastic noise, compression artifacts, and the possible mismatch between the forward observation model and the camera device. The operator $\bH$ is usually ill-conditioned or singular due to the presence of unknown noise ($\eta$) that makes the SISR a highly ill-posed nature of inverse problems. Since, due to the ill-posed nature, there are many possible solutions, regularization is required to select the most plausible ones.

\begin{figure}[t]
	\begin{center}
		\includegraphics[width=\linewidth]{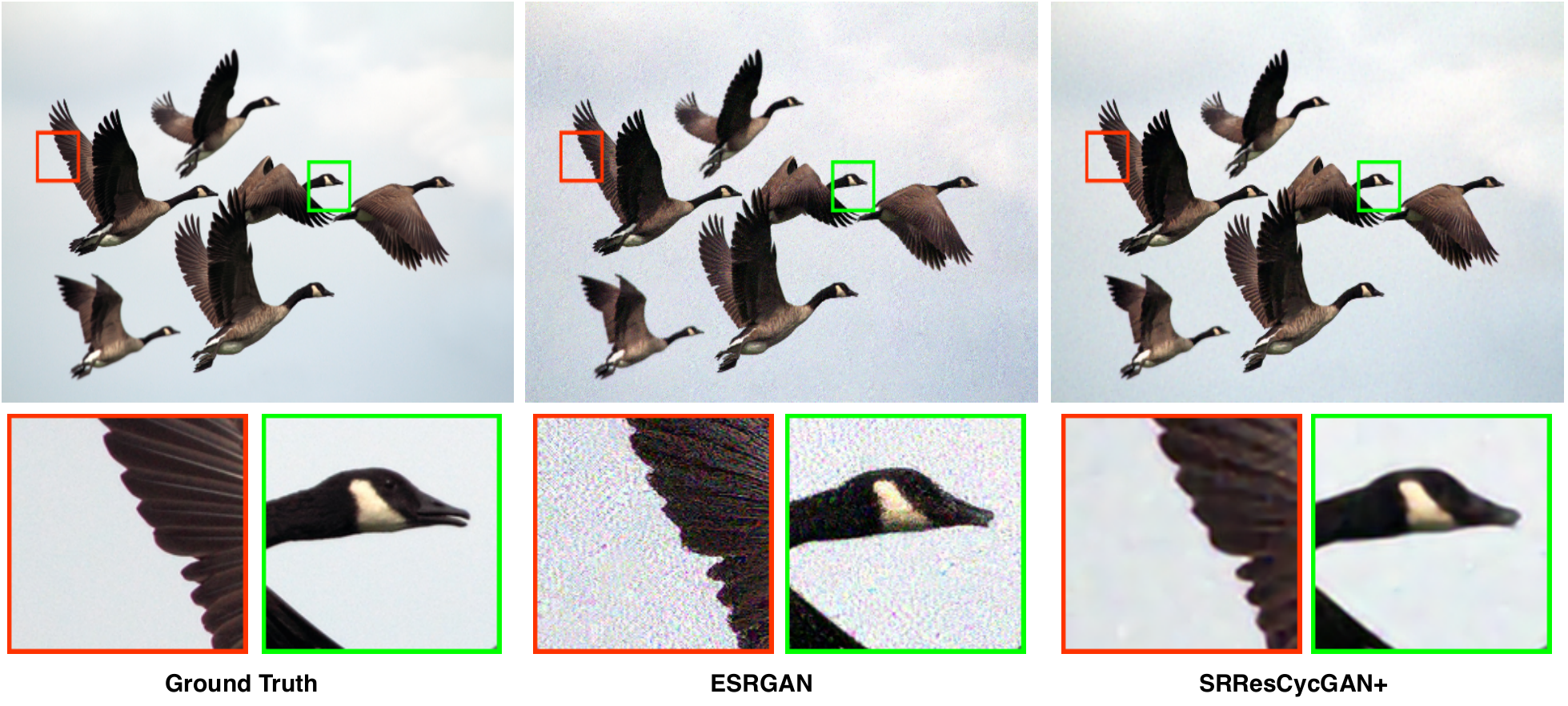}
	\end{center}
	\vspace{-0.4cm}
	\caption{The super-resolution results at the $\times 4$ upscaling factor of the state-of-art--ESRGAN, the proposed SRResCycGAN+ with respect to the ground-truth images. SRResCycGAN+ has successfully remove the visible artifacts, while the ESRGAN has still artifacts due to data bias between the training and testing images.}
	\vspace{-0.5cm}
	\label{fig:tease}
\end{figure}
Recently, numerous works have been addressed on the task of SISR that are based on deep CNNs for their powerful feature representation capabilities either on PSNR values~\cite{kim2016vdsrcvpr,Lim2017edsrcvprw,kai2017ircnncvpr,kai2018srmdcvpr,yuan2018unsupervised,Li2019srfbncvpr,zhang2019deep} or on visual quality~\cite{ledig2017srgan,wang2018esrgan}. These SR methods mostly rely on the known degradation operators such as bicubic (\ie noise-free) with paired LR and HR images (same clean domain) in the supervised training, while other methods do not follow the image observation (physical) model (refers to Eq.~\eqref{eq:degradation_model}). In the real-world settings, the input LR images suffer from different kinds of degradation or LR is different from the HR domain. Under such circumstances, these SR methods often fail to produce convincing SR results. In Figure~\ref{fig:tease}, we show the results of the state-of-art deep learning method--ESRGAN with the noisy input image. The ESRGAN degraded SR result is due to the difference of training and testing data domains. The detailed analysis of the deep learning-based SR models on the real-world data can be found in the recent literature~\cite{lugmayr2019unsupervised,fritsche2019dsgan}.

\begin{figure}[t]
\centering
\includegraphics[scale=1.0]{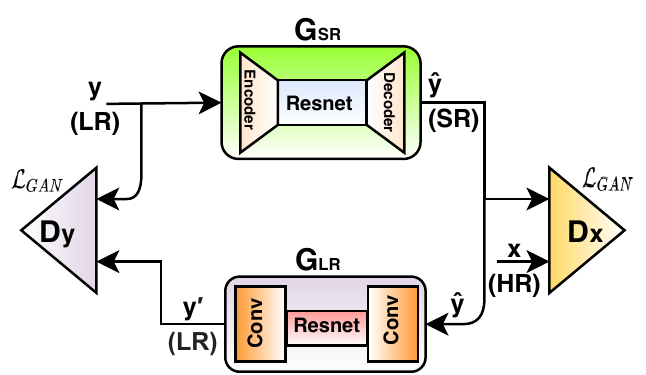}
\caption{Visualizes the structure of the our proposed SR approach setup. We trained the network $\mathbf{G}_{SR}$ in a GAN framework, where our goal is to map images from the LR ($\by$) to the HR ($\bx$), while maintaining the domain consistency between the LR and HR images.}
\label{fig:srrescycgan}
\vspace{-0.5cm}
\end{figure}

In this work, we propose a SR learning method (SRResCycGAN) that overcomes the challenges of real image super-resolution. It is inspired by CycleGAN~\cite{zhu2017unpairedcycgan} structure which maintains the domain consistency between the LR and HR domain. It is also inspired by powerful image regularization and large-scale optimization techniques to solve general inverse problems in the past. The scheme of our proposed real image SR approach setup is shown in Fig.~\ref{fig:srrescycgan}. The $\mathbf{G}_{SR}$ network takes the input LR image and produces the SR output with the supervision of the SR discriminator network $\mathbf{D}_{\bx}$. For the domain consistency between the LR and HR, the $\mathbf{G}_{LR}$ network reconstructs the LR image from the SR output with the supervision of the LR discriminator network $\mathbf{D}_{\by}$.

We evaluate our proposed SR method on multiple datasets with synthetic and natural image corruptions. We use the Real-World Super-resolution (RWSR) dataset~\cite{NTIRE2020RWSRchallenge} to show the effectiveness of our method through quantitative and qualitative experiments. Finally, we also participated in the AIM2020 Real Image Super-resolution Challenge~\cite{AIM2020_RSRchallenge} for the Track-3 ($\times 4$ upscaling) associated with the ECCV 2020 workshops. Table~\ref{tab:track3} shows the final testset SR results for the track-3 of our method (\textbf{MLP\_SR}) with others as well as the visual comparison in the Fig.~\ref{fig:4x_result_val} and Fig.~\ref{fig:4x_result_test}. 
\section{Related Work}
\subsection{Image Super-Resolution methods}
Recently, numerous works have addressed the task of SISR using deep CNNs for their powerful feature representation capabilities. A preliminary CNN-based method to solve SISR is a super-resolution convolutional network with three layers (SRCNN)~\cite{dong2014srcnneccv}. Kim~\etal\cite{kim2016vdsrcvpr} proposed a very deep SR (VDSR) network with residual learning approach. The efficient subpixel convolutional network (ESPCNN)~\cite{Shi2016pixelcnncvpr} was proposed to take bicubicly LR input and introduced an efficient subpixel convolution layer to upscale the LR feature maps to HR images at the end of the network. Lim \etal\cite{Lim2017edsrcvprw} proposed an enhanced deep SR (EDSR) network by taking advantage of the residual learning. Zhang~\etal\cite{kai2017ircnncvpr} proposed iterative residual convolutional network (IRCNN) to solve SISR problem by using a plug-and-play framework. Zhang~\etal\cite{kai2018srmdcvpr} proposed a deep CNN-based super-resolution with multiple degradation (SRMD). Yaoman \etal\cite{Li2019srfbncvpr} proposed a feedback network (SRFBN) based on feedback connections and recurrent neural network-like structure. Zhang \etal\cite{zhang2019deep} proposed a deep plug-and-play Super-Resolution method for arbitrary blur kernels by following the multiple degradation.  In \cite{srwdnet}, the authors proposed SRWDNet to solve the joint deblurring and super-resolution task by following the realistic degradation. These methods mostly rely on the PSNR-based metric by optimizing the $\mathcal{L}_1$/$\mathcal{L}_2$ losses with blurry results in a supervised way, while they do not preserve the visual quality with respect to human perception. Moreover, the above-mentioned methods are deeper or wider CNN networks to learn non-linear mapping from LR to HR with a large number of training samples, while neglecting the real-world settings.

\subsection{ Real Image Super-Resolution methods}
\begin{figure}[t]
\centering
\includegraphics[width=\linewidth]{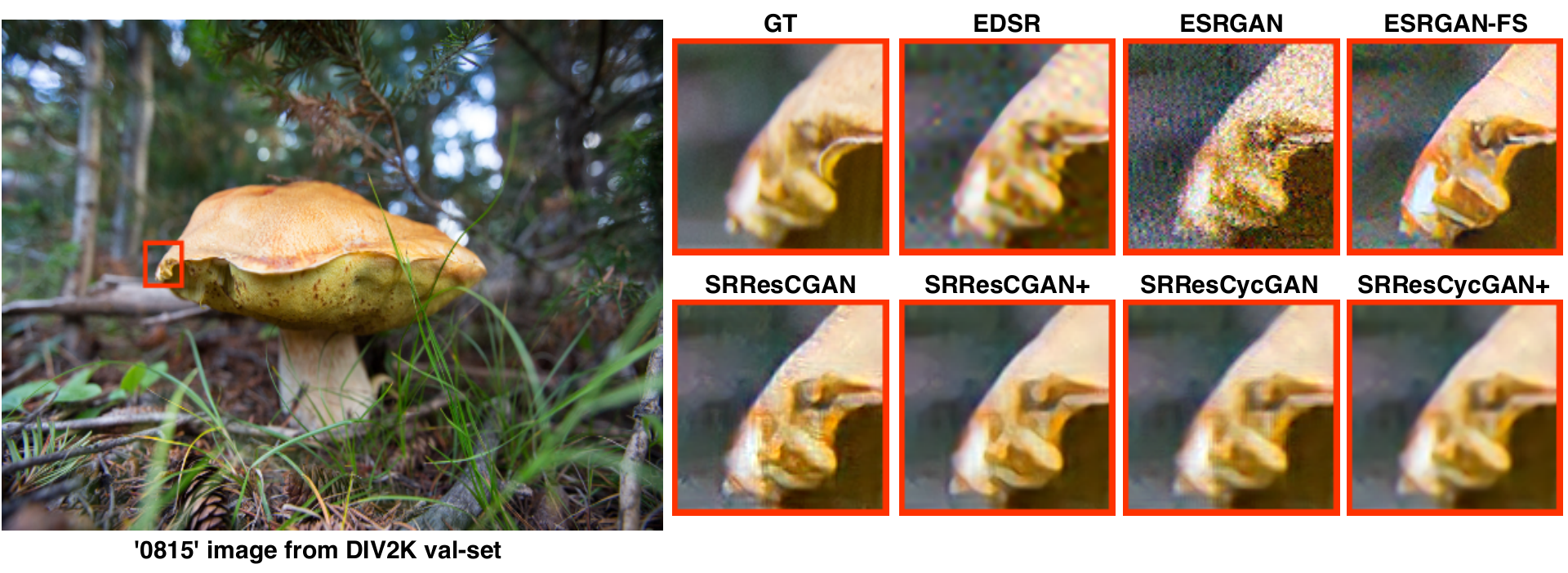}
\vspace{-0.4cm}
\caption{Visual comparison of our method with the other state-of-art methods on the DIV2K validation set at the $\times4$ super-resolution.}
\label{fig:4x_result_div2k}
\vspace{-0.5cm}
\end{figure}

For the perception SR task, a preliminary attempt was made by Ledig \etal\cite{ledig2017srgan} who proposed the SRGAN method to produce perceptually more pleasant results. To further enhance the performance of the SRGAN, Wang \etal\cite{wang2018esrgan} proposed the ESRGAN model to achieve the state-of-art perceptual performance. Despite their success, the previously mentioned methods are trained with HR/LR image pairs on the bicubic down-sampling \ie noise-free and thus they have limited performance in the real-world settings. More recently, Lugmayr \etal\cite{lugmayr2019unsupervised} proposed a benchmark protocol for the real-wold image corruptions and introduced the real-world challenge series~\cite{AIM2019RWSRchallenge} that described the effects of bicubic downsampling and separate degradation learning for super-resolution. Later on, Fritsche \etal\cite{fritsche2019dsgan} proposed the DSGAN to learn degradation by training the network in an unsupervised way and modified the ESRGAN structure as the ESRGAN-FS to further enhance the performance in the real-world settings. Recently, the authors proposed the SRResCGAN~\cite{muhammad2020srrescgan} to solve real-world SR problem, which is inspired by a physical image formation model. However, the above methods still suffer unpleasant artifacts (see the Fig.~\ref{fig:4x_result_div2k} and the Table~\ref{tab:comp_sota}). Our approach takes into account the real-world settings by greatly increasing its applicability in practical scenarios. 

\section{Proposed Method}
\label{sec:proposed_method}
\subsection{Problem Formulation}
By referencing to the Eq.~\eqref{eq:degradation_model}, the recovery of $\bx$ from $\by$ mostly relies on the variational approach for combining the observation and prior knowledge, and is given by the following objective function:
\begin{equation}
    \mathbf{J}(\bx) = \underset{\mathbf{x}}{\arg \min }~\frac{1}{2}\|\by - \bH*\bx\|_2^{2}+\lambda \mathcal{R}(\bx),
    \label{eq:eq1}
\end{equation}
where $\frac{1}{2}\|\by-\mathbf{H}*\bx\|_2^2$ is the data fidelity (also known as log-likelihood) term that measures the proximity of the solution to the observations, $\mathcal{R}(\bx)$ is the regularization term that is associated with image priors, and $\lambda$ is the trade-off parameter that governs the compromise between the data fidelity and the regularizer term. Interestingly, the variational approach has a direct link to the Bayesian approach and the derived solutions can be described either as penalized maximum likelihood or as maximum a posteriori (MAP) estimates~\cite{bertero1998map1,figueiredo2007map2}. Thanks to the recent advances of deep learning, the regularizer (\ie $\mathcal{R}(\bx)$) is employed by the SRResCGAN~\cite{muhammad2020srrescgan} generator structure that has powerful image priors capabilities.

\subsection{SR Learning Model}
\label{sec:sr_learning}
The proposed Real Image SR approach setup is shown in the Fig.~\ref{fig:srrescycgan}. The SR generator network $\mathbf{G}_{SR}$ borrowed from the SRResCGAN~\cite{muhammad2020srrescgan} is trained in a GAN~\cite{goodfellow2014gan} framework by using the LR ($\by$) images with their corresponding HR images with pixel-wise supervision in the clean HR target domain ($\bx$), while maintaining the domain consistency between the LR and HR images. In the next coming sections~\ref{sec:net_arch}, \ref{sec:net_losses}, and \ref{sec:net_training}, we present the details of the network architectures, network losses, and training descriptions for the proposed SR setup. 

\subsection{Network Architectures}
\label{sec:net_arch}
\subsubsection{SR Generator ($\mathbf{G_{SR}}$):}
We use the SR generator $\mathbf{G_{SR}}$ network which is basically an \emph{Encoder-Resnet-Decoder} like structure as done SRResCGAN~\cite{muhammad2020srrescgan}. In the $\mathbf{G_{SR}}$ network, both \emph{Encoder} and \emph{Decoder} layers have $64$ convolutional feature maps of $5\times5$ kernel size with $C \times H\times W$ tensors, where $C$ is the number of channels of the input image. Inside the \emph{Encoder}, LR image is upsampled by the Bicubic kernel with \emph{Upsample} layer, where the choice of the upsampling kernel is arbitrary. \emph{Resnet} consists of $5$ residual blocks with two Pre-activation \emph{Conv} layers, each of $64$ feature maps with kernel support $3\times3$, and the pre-activation is the parametrized rectified linear unit (PReLU) with $64$ output feature channels. The trainable projection layer~\cite{Lefkimmiatis2018UDNet} inside the \emph{Decoder} computes the proximal map with the estimated noise standard deviation $\sigma$ and handles the data fidelity and prior terms. The noise realization is estimated in the intermediate \emph{Resnet} that is sandwiched between \emph{Encoder} and \emph{Decoder}. The estimated residual image after \emph{Decoder} is subtracted from the LR input image. Finally, the clipping layer incorporates our prior knowledge about the valid range of image intensities and enforces the pixel values of the reconstructed image to lie in the range $[0, 255]$. The reflection padding is also used before all the \emph{Conv} layers to ensure slowly varying changes at the boundaries of the input images.

\subsubsection{SR Discriminator ($\mathbf{D}_{\bx}$):}
The SR discriminator network is trained to discriminate the real HR images from the fake HR images generated by the $\mathbf{G_{SR}}$. The raw discriminator network contains 10 convolutional layers with kernels support $3\times3$ and $4\times4$ of increasing feature maps from $64$ to $512$ followed by Batch Norm (BN) and leaky ReLU as do in SRGAN~\cite{ledig2017srgan}. 

\subsubsection{LR Generator ($\mathbf{G_{LR}}$):}
We adapt the similar architecture as does in \cite{yuan2018unsupervised} for the down-sampling which is basically a \emph{Conv-Resnet-Conv} like structure. We use 6 residual blocks in the \emph{Resnet} with 3 convolutional layers at the head and tail \emph{Conv}, while the stride is set to 2 in the second and third head \emph{Conv} layers for the down-sampling purpose. 

\subsubsection{LR Discriminator ($\mathbf{D}_{\by}$):}
The LR discriminator network consists of a three-layer convolutional network that operates on the patch level as do in PatchGAN~\cite{isola2017image,li2016precomputed}. All the \emph{Conv} layers have $5\times5$ kernel support with feature maps from 64 to 256 and also applied the Batch Norm and Leaky ReLU (LReLU) activation after each \emph{Conv} layer except the last \emph{Conv} layer that maps 256 to 1 features.

\subsection{Network Losses}
\label{sec:net_losses}
To learn the image super-resolution, we train the proposed SRResCycGAN network with the following loss functions:
\begin{equation}
    \mathcal{L}_{G_{SR}} = \mathcal{L}_{\mathrm{per}}+ \mathcal{L}_{\mathrm{GAN}} + \mathcal{L}_{tv} + 10\cdot \mathcal{L}_{\mathrm{1}} +  10\cdot \mathcal{L}_{\mathrm{cyc}}
    \label{eq:l_g}
\end{equation}
where, these losses are defined as follows:\\
\textbf{Perceptual loss ($\mathcal{L}_{\mathrm{per}}$):} It focuses on the perceptual quality of the output image and is defined as:
\begin{equation}
    \mathcal{L}_{\mathrm{per}}=\frac{1}{N} \sum_{i}^{N}\mathcal{L}_{\mathrm{VGG}}=\frac{1}{N} \sum_{i}^{N}\|\phi(\mathbf{G}_{SR}(\by_i))-\phi(\bx_i)\|_{1}
\end{equation}
where, $\phi$ is the feature extracted from the pretrained VGG-19 network at the same depth as ESRGAN~\cite{wang2018esrgan}.\\
\textbf{Texture loss ($\mathcal{L}_{\mathrm{GAN}}$):} It focuses on the high frequencies of the output image and it is defined as:
\begin{equation}
   \mathbf{D}_{\bx}(\bx, \hat{\by})(C) = \sigma(C(\bx)-\mathbb{E}[C(\hat{\by})])
\end{equation}
Here, $C$ is the raw discriminator output and $\sigma$ is the sigmoid function. By using the relativistic discriminator~\cite{wang2018esrgan}, we have:
\begin{equation}
    \begin{split}
       \mathcal{L}_{\mathrm{GAN}} = \mathcal{L}_{\mathrm{RaGAN}} = &-\mathbb{E}_{\bx}\left[\log \left(1-\mathbf{D}_{\bx}(\bx, \mathbf{G}_{SR}(\by))\right)\right] \\
    &-\mathbb{E}_{\hat{\by}}\left[\log \left(\mathbf{D}_{\bx}(\mathbf{G}_{SR}(\by), \bx)\right)\right] 
    \end{split}
\end{equation}
where, $\mathbb{E}_{\bx}$ and $\mathbb{E}_{\hat{\by}}$ represent the operations of taking average for all real ($\bx$) and fake ($\hat{\by}$) data in the mini-batches respectively. \\
\textbf{Content loss ($\mathcal{L}_{\mathrm{1}}$):} It is defined as:
\begin{equation}
    \mathcal{L}_{1} = \frac{1}{N} \sum_{i}^{N} \|\mathbf{G}_{SR}(\by_i)-\bx_i\|_{1}
\end{equation}
where, $N$ represents the size of mini-batch.\\
\textbf{TV (total-variation) loss ($\mathcal{L}_{tv}$):} It focuses to minimize the gradient discrepancy and produces sharpness in the output SR image and it is defined as:
\begin{equation}
   \mathcal{L}_{tv}=\frac{1}{N} \sum_{i}^{N}\left(\left\|\nabla_{h} \mathbf{G}_{SR}\left(\by_{i}\right) - \nabla_{h} \left(\bx_{i}\right) \right\|_{1}+\left\|\nabla_{v} \mathbf{G}_{SR}\left(\by_{i}\right) - \nabla_{v} \left(\bx_{i}\right) \right\|_{1}\right)
\end{equation}
Here, $\nabla_{h}$ and $\nabla_{v}$ denote the horizontal and vertical gradients of the images.\\
\textbf{Cyclic loss ($\mathcal{L}_{\mathrm{cyc}}$):} It focuses to maintain the cyclic consistency between LR and HR domain and it is defined as:
\begin{equation}
    \mathcal{L}_{cyc} = \frac{1}{N} \sum_{i}^{N} \|\mathbf{G}_{LR}(\mathbf{G}_{SR}(\by_i))-\by_i\|_{1}
\end{equation}

\subsection{Training description}
\label{sec:net_training}
At the training phase, we set the input LR patches size as $32\times32$ with their corresponding HR patches. We train the network in an end-to-end manner for 51000 training iterations with a batch size of 16 using Adam optimizer with parameters $\beta_1 =0.9$, $\beta_2=0.999$, and $\epsilon=10^{-8}$ without weight decay for generators ($\mathbf{G_{SR}}$ \& $\mathbf{G_{LR}}$) and discriminators ($\mathbf{D}_{\bx}$ \& $\mathbf{D}_{\by}$) to minimize the loss in Eq.~\eqref{eq:l_g}. The learning rate is initially set to $10^{-4}$ and then multiplies by $0.5$ after 5K, 10K, 20K, and 30K iterations. The projection layer parameter $\sigma$ is estimated according to \cite{liu2013single} from the input LR image. 

\section{Experiments}
\subsection{Training data}
We use the source domain data ($\Tilde{\by}$: 2650 HR images) that are corrupted with two known degradation, e.g., sensor noise, compression artifacts as well as unknown degradation, and target domain data ($\bx$: 800 clean HR images from the DIV2K~\cite{div2k}) provided in the NTIRE2020 Real-World Super-resolution (RWSR)  Challenge~\cite{NTIRE2020RWSRchallenge} for the track-1. We use the source and target domain data for training the $\mathbf{G_{SR}}$ network under the different degradation scenarios. The LR data ($\by$) with similar corruption as in the source domain is generated from the down-sample GAN network (DSGAN)~\cite{fritsche2019dsgan} with their corresponding HR target domain ($\bx$) images. Furthermore, we use the training data (\ie $\by$: 19000 LR images, $\bx$: 19000 HR images) provided in the AIM2020 Real Image SR Challenge~\cite{AIM2020_RSRchallenge} for the track-3 ($\times 4$ upscaling) for training the SRResCycGAN (refer to the section-\ref{sec:aim2020_risr}).  
\begin{table}[t]
	\centering
	\vspace{-0.5cm}
	\caption{The $\times4$ SR quantitative results comparison of our method with others over the DIV2K validation-set (100 images). Top section: SR results comparison with added sensor noise ($\sigma=8$) and compression artifacts ($quality=30$) in the validation-set. Middle section: SR results with the unknown corruptions (e.g., sensor noise, compression artifacts, etc.) in the validation-set provided in the RWSR challenge series~\cite{AIM2019RWSRchallenge,NTIRE2020RWSRchallenge}. 
    Bottom section: SR results with the real image corruptions in the validation-set and testset provided in the AIM 2020 Real Image SR challenge~\cite{AIM2020_RSRchallenge} for the track-3. The arrows indicate if high $\uparrow$ or low $\downarrow$ values are desired. The best performance is shown in {\color{red} red} and the second best performance is shown in {\color{blue} blue}.}
    \vspace{0.1cm}
	\tabcolsep=0.01\linewidth
	\scriptsize
	\resizebox{1.0\textwidth}{!}{
	\begin{tabular}{|c|c|c|c|c|c|c|c|}
	    \multicolumn{2}{c}{ } & \multicolumn{3}{c}{sensor noise ($\sigma=8$)} & \multicolumn{3}{c}{compression artifacts ($q=30$)} \\
		 SR methods & \#Params & PSNR$\uparrow$ & SSIM$\uparrow$ & LPIPS$\downarrow$ & PSNR$\uparrow$ & SSIM$\uparrow$ & LPIPS$\downarrow$  \\ 
		\hline
		EDSR~\cite{Lim2017edsrcvprw} & $43M$ & 24.48 & 0.53 & 0.6800 & 23.75  & 0.62 & 0.5400 \\ 
		ESRGAN~\cite{wang2018esrgan} & $16.7M$ & 17.39 & 0.19 & 0.9400 & 22.43   & 0.58 & 0.5300 \\
	    ESRGAN-FT~\cite{lugmayr2019unsupervised} & $16.7M$ & 22.42 & 0.55 & 0.3645 & 22.80  & 0.57  & {\color{red}0.3729} \\
		ESRGAN-FS~\cite{fritsche2019dsgan} & $16.7M$ & 22.52 & 0.52 & {\color{red}0.3300} & 20.39 & 0.50 & {\color{blue}0.4200} \\
		SRResCGAN~\cite{muhammad2020srrescgan} & $380K$ & 25.46 & 0.67 & {\color{blue}0.3604}  & 23.34  & 0.59 & 0.4431 \\
		SRResCycGAN (ours) & $380K$ & {\color{blue}25.98} & {\color{blue}0.70} & 0.4167  & {\color{blue}23.96} & {\color{blue}0.63} & 0.4841 \\
		SRResCycGAN+ (ours) & $380K$ & {\color{red}26.27} & {\color{red}0.72} & 0.4542  & {\color{red}24.05} & {\color{red}0.64} & 0.5192 \\
		\hline
		&\multicolumn{3}{c}{ } & \multicolumn{3}{c}{unknown corruptions~\cite{NTIRE2020RWSRchallenge}}  &\\
		 \hline
		SRResCGAN~\cite{muhammad2020srrescgan} & $380K$ & 25.05 & 0.67 & {\color{red}0.3357} \\
		SRResCycGAN (ours) & $380K$ & {\color{blue}26.13} & {\color{blue}0.71} & {\color{blue}0.3911} \\
		SRResCycGAN+ (ours) & $380K$ & {\color{red}26.39} & {\color{red}0.73} & 0.4245 \\
		\hline
		&\multicolumn{3}{c}{ } & \multicolumn{3}{c}{real image corruptions~\cite{AIM2020_RSRchallenge}}  &\\
		 \hline
		SRResCycGAN (ours, valset) & $380K$ & 28.6239 & 0.8250 & - \\
		SRResCycGAN (ours, testset) & $380K$ & 28.6185 & 0.8314 & - \\
	\end{tabular}}
	\label{tab:comp_sota}
	\vspace{-0.3cm}
\end{table}

\subsection{Technical details}
We implemented our method in the Pytorch. The experiments are performed under Windows 10 with i7-8750H CPU with 16GB RAM and on the NVIDIA RTX-2070 GPU with 8GB memory. It takes about 25 hours to train the network. The run time per image (on the GPU) is  4.54 seconds at the AIM2020 Real Image SR testset. In order to further enhance the fidelity, we use a self-ensemble strategy~\cite{timofte2016seven} (denoted as SRResCycGAN+) at the test time, where the LR inputs are flipped/rotated and the SR results are aligned and averaged for enhanced prediction.

\subsection{Evaluation metrics}
We evaluate the trained model under the Peak Signal-to-Noise Ratio (PSNR), Structural Similarity (SSIM), and LPIPS~\cite{zhang2018unreasonable} metrics. The PSNR and SSIM are distortion-based measures that correlate poorly with actual perceived similarity, while LPIPS better correlates with human perception than the distortion-based/handcrafted measures. As LPIPS is based on the features of pretrained neural networks, so we use it for the quantitative evaluation with features of AlexNet~\cite{zhang2018unreasonable}. The quantitative SR results are evaluated on the $RGB$ color space.
\subsection{Comparison with the state-of-art methods}
\label{sec:comp_sota}
We compare our method with other state-of-art SR methods including EDSR~\cite{Lim2017edsrcvprw}, ESRGAN~\cite{wang2018esrgan}, ESRGAN-FT~\cite{lugmayr2019unsupervised}, ESRGAN-FS~\cite{fritsche2019dsgan}, and SRResCGAN~\cite{muhammad2020srrescgan}, whose source codes are available online. The two degradation settings (\ie sensor noise, JPEG compression) have been considered under the same experimental situations for all methods. We run all the original source codes and trained models by the default parameters settings for the comparison. The EDSR is trained without the perceptual loss (only $\mathcal{L}_{\mathrm{1}}$) by a deep SR residual network using the bicubic supervision. The ESRGAN is trained with the $\mathcal{L}_{\mathrm{perceptual}}$, $\mathcal{L}_{\mathrm{GAN}}$, and  $\mathcal{L}_{\mathrm{1}}$ by a deep SR network using the bicubic supervision. The ESRGAN-FT and ESRGAN-FS apply the same SR architecture and perceptual losses as in the ESRGAN using the two known degradation supervision. The SRResCGAN is trained with the similar losses combination as done in the ESRGAN using the two known degradation supervision. We train the proposed SRResCycGAN with the similar losses combination as done in the ESRGAN and SRResCGAN with the additional cyclic loss by using the bicubic supervision.  

\begin{figure}[htbp!]
    \centering
    \begin{subfigure}[t]{0.85\textwidth}
        \centering
        \includegraphics[width=\linewidth]{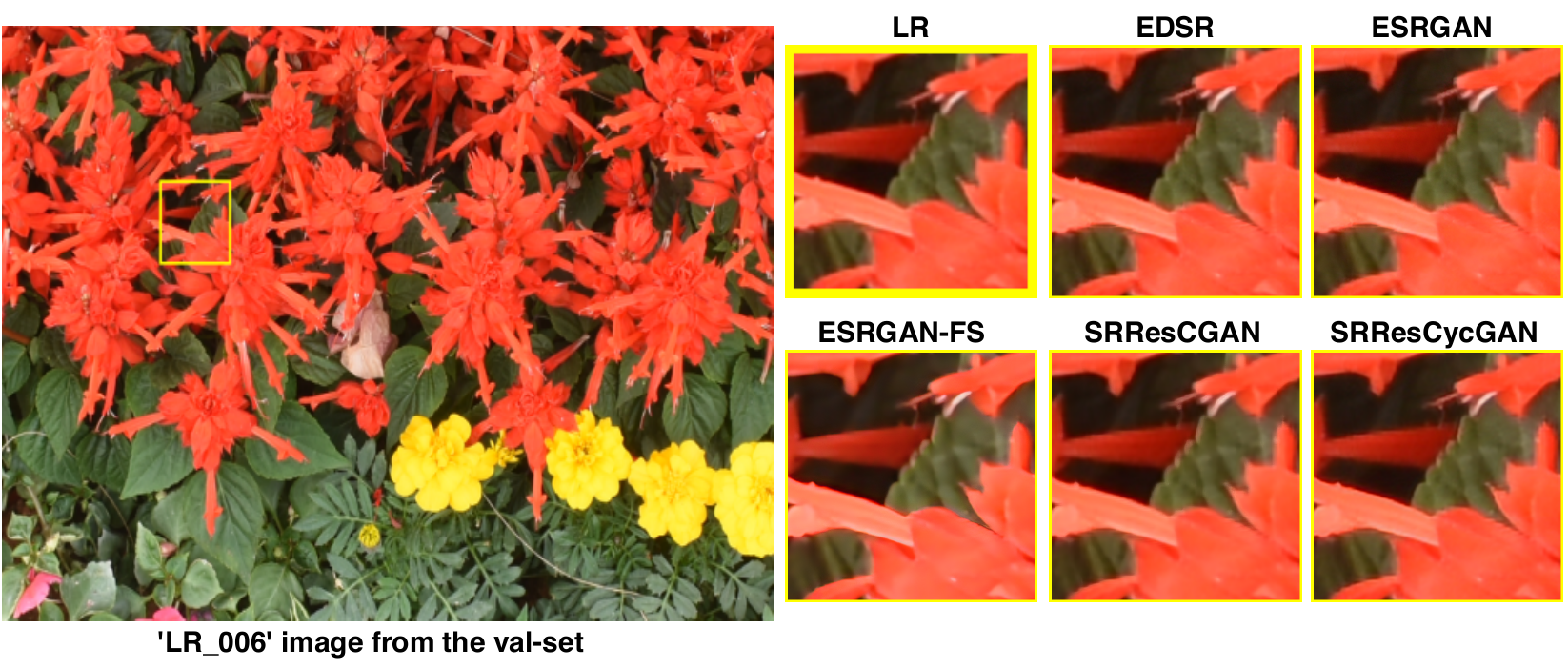}
    \end{subfigure}\\ 
    \begin{subfigure}[t]{0.85\textwidth}
        \centering
        \includegraphics[width=\linewidth]{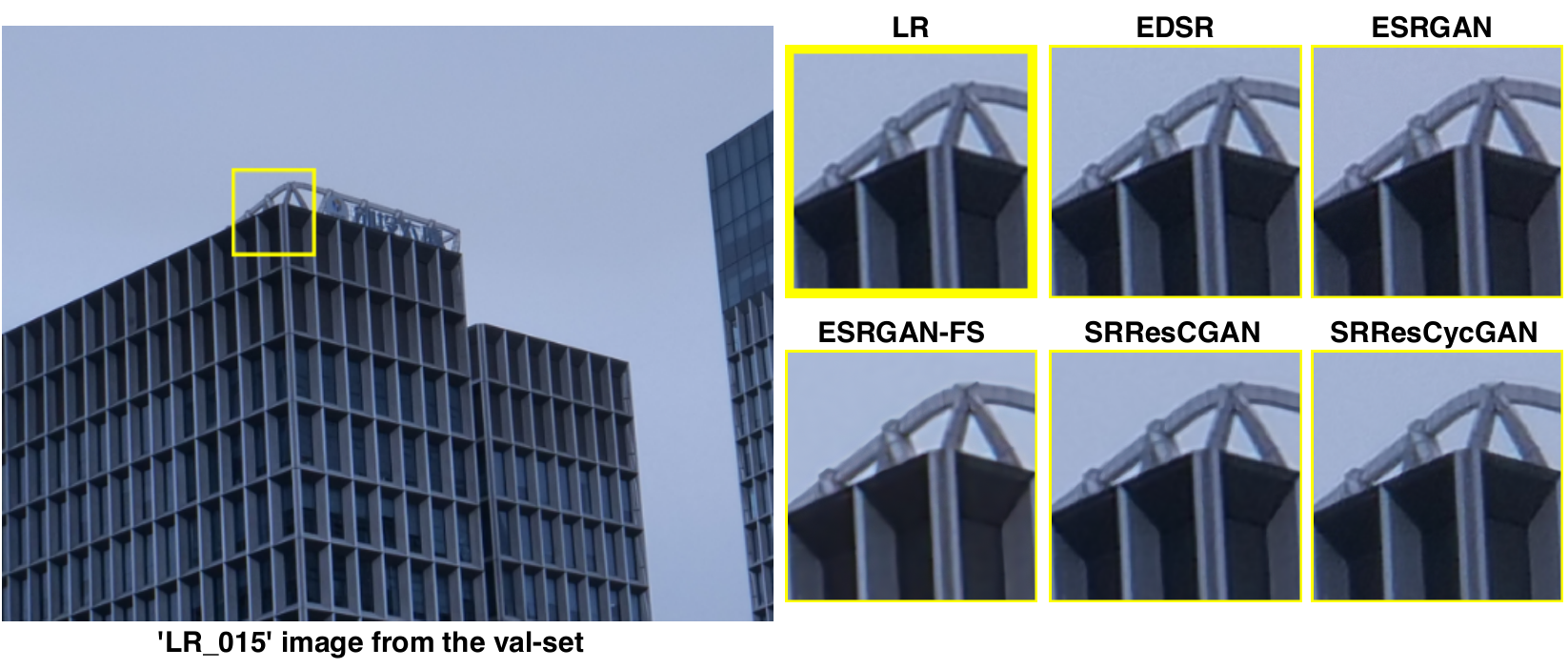}
    \end{subfigure}
    \vspace{-0.4cm}
    \caption{Visual comparison of our method with the other state-of-art methods on the AIM 2020 Real Image SR (track-3) validation set at the $\times4$ super-resolution.}
    \label{fig:4x_result_val}
    \vspace{-0.5cm}
\end{figure}

Table~\ref{tab:comp_sota} shows the quantitative results comparison of our method over the DIV2K validation-set (100 images) with two known degradation (\ie sensor noise, JPEG compression), the unknown degradation in the NTIRE2020 Real-World SR challenge series~\cite{NTIRE2020RWSRchallenge}, and the validation-set and testset in the AIM2020 Real Image SR Challenge~\cite{AIM2020_RSRchallenge}. Our method results outperform in terms of PSNR and SSIM compared to the other methods, while in the case of LPIPS, we have comparable results with others. In the case of the sensor noise ($\sigma=8$) and JPEG compression ($q=30$) in the top section of the Table~\ref{tab:comp_sota}, the ESRGAN has the worst performance in terms of the PSNR, SSIM, and LPIPS among all methods. Its also depicts the visual quality in Fig.~\ref{fig:4x_result_val}. The EDSR has better performance to the noisy input, but it produces more blurry results. These are due to the domain distribution difference by the bicubic down-sampling during training phase. The ESRGAN-FT and ESRGAN-FS have much better performance due to overcoming the domain distribution shift problem, but they have still visible artifacts. The SRResCGAN has better robustness to the noisy input, but still has lower the PSNR and SSIM due to lacking the domain consistency problem. The proposed method has successfully overcome the challenge of the domain distribution shift in both degradation settings, which depicts in the both quantitative and qualitative results. In the middle section of the Table~\ref{tab:comp_sota},for the unknown degradation in the NTIRE2020 Real-World SR challenge~\cite{NTIRE2020RWSRchallenge}, the SRResCycGAN has much better the PSNR/SSIM improvment, while the LPIPS is also comparable with the SRResCGAN. In the bottom section of the Table~\ref{tab:comp_sota}, we also report the validation-set and testset SR results in the AIM2020 Real Image SR Challenge~\cite{AIM2020_RSRchallenge} for the track-3. Despite that, the parameters of the proposed $\mathbf{G}_{SR}$ network are much less, which makes it suitable for deployment in mobile/embedded devices where memory storage and CPU power are limited as well as good image reconstruction quality.

Regarding the visual quality, Fig.~\ref{fig:4x_result_val} shows the qualitative comparison of our method with other SR methods at the $\times 4$ upscaling factor on the validation-set~\cite{NTIRE2020RWSRchallenge}. In contrast to the existing state-of-art methods, our proposed method produces the excellent SR results that are reflected in the PSNR/SSIM values, as well as the visual quality of the reconstructed images with almost no visible corruptions.
\begin{table}[htbp!]
    \vspace{-0.5cm}
	\centering%
	\caption{Final Testset results for the Real Image SR ($\times4$)  challenge Track-3~\cite{AIM2020_RSRchallenge}. The table contains ours (\textbf{MLP\_SR}) with other methods that are ranked in the challenge. The participating methods are ranked according to their weighted Score of the PSNR and SSIM given in the AIM 2020 Real Image SR Challenge~\cite{AIM2020_RSRchallenge}.}
	\vspace{0.1cm}
	\resizebox{0.7\textwidth}{!}{%
		\begin{tabular}{|l|c|c|c|}
                     Team Name &        PSNR$\uparrow$ &       SSIM$\uparrow$  & Weighed\_score$\uparrow$ \\
        \hline
         Baidu &  $31.3960$ &  $0.8751$ & $0.7099_{(1)}$\\
         ALONG &  $31.2369$ &  $0.8742$ & $0.7076_{(2)}$\\
         CETC-CSKT &  $31.1226$ &  $0.8744$ & $0.7066_{(3)}$\\
         SR-IM &  $31.1735$ &  $0.8728$ & $0.7057$\\
         DeepBlueAI &  $30.9638$ &  $0.8737$ & $0.7044$\\
         JNSR &  $30.9988$ &  $0.8722$ & $0.7035$\\
         OPPO\_CAMERA &  $30.8603$ &  $0.8736$ & $0.7033$\\
         Kailos &  $30.8659$ &  $0.8734$ & $0.7031$\\
         SR\_DL &  $30.6045$ &  $0.8660$  & $0.6944$\\
         Noah\_TerminalVision &  $30.5870$ &  $0.8662$ & $0.6944$\\
         Webbzhou &  $30.4174$ &  $0.8673$ & $0.6936$\\
         TeamInception &  $30.3465$ &  $0.8681$ & $0.6935$\\
         IyI &  $30.3191$ &  $0.8655$ & $0.6911$\\
         MCML-Yonsei &  $30.4201$ &  $0.8637$ & $0.6906$\\
         MoonCloud &  $30.2827$ &  $0.8644$ & $0.6898$\\
         qwq &  $29.5878$ &  $0.8547$ & $0.6748$\\
         SrDance &  $29.5952$ &  $0.8523$ & $0.6729$\\
         \textbf{MLP\_SR} &  $28.6185$ &  $0.8314$ & $0.6457$\\
         RRDN\_IITKGP &  $27.9708$ &  $0.8085$ & $0.6201$\\
         congxiaofeng &  $26.3915$ &  $0.8258$ & $0.6187$\\
        \end{tabular}
		}
	\label{tab:track3}
	\vspace{-0.5cm}
\end{table}

\begin{figure}[htbp!]
    \centering
    \begin{subfigure}[t]{0.85\textwidth}
        \centering
        \includegraphics[width=\linewidth]{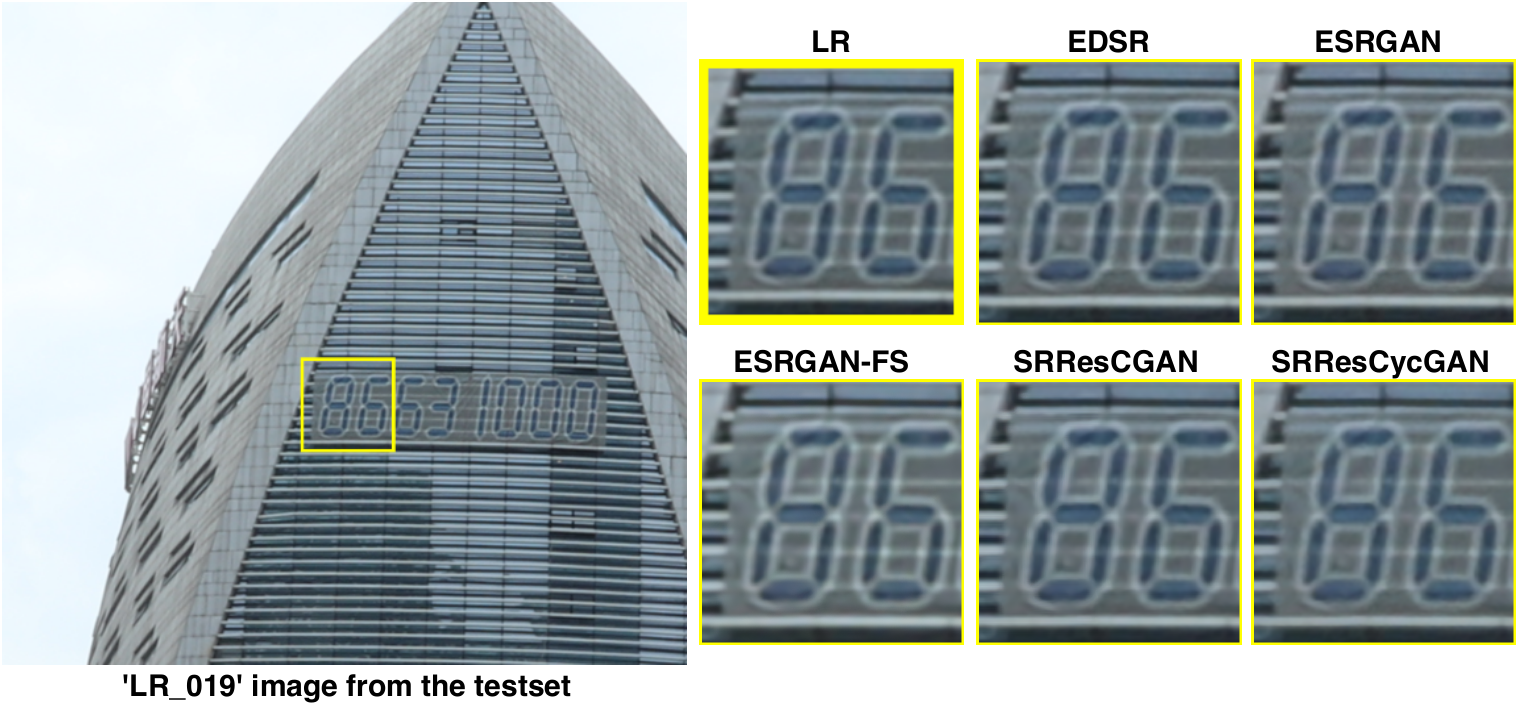}
    \end{subfigure}\\ 
    \begin{subfigure}[t]{0.85\textwidth}
        \centering
        \includegraphics[width=\linewidth]{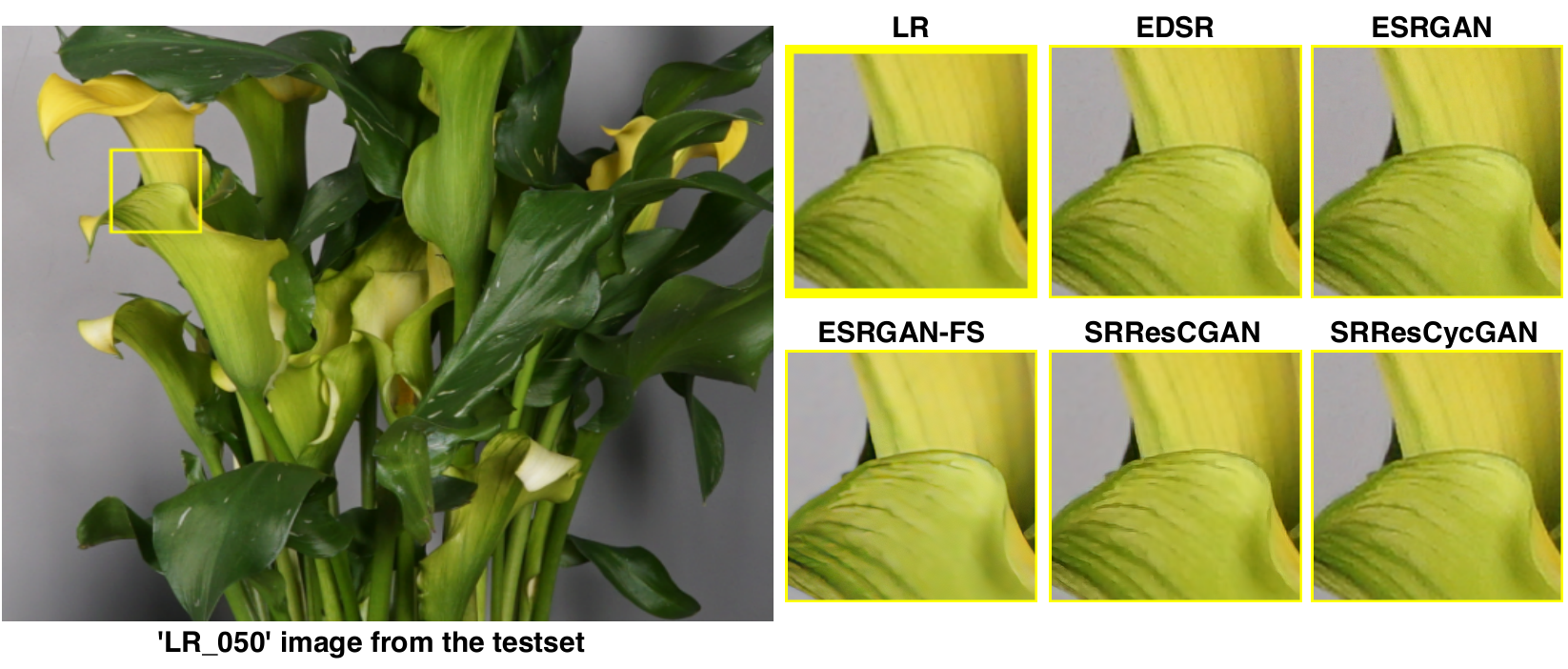}
    \end{subfigure}
    \vspace{-0.4cm}
    \caption{Visual comparison of our method with the other state-of-art methods on the AIM 2020 Real Image SR (track-3) test set at the $\times4$ super-resolution.}
    \label{fig:4x_result_test}
    \vspace{-0.5cm}
\end{figure}
\subsection{The AIM 2020 Real Image SR Challenge ($\times4$)}
\label{sec:aim2020_risr}
We participated in the AIM2020 Real Image Super-Resolution Challenge~\cite{AIM2020_RSRchallenge} for the track-3 ($\times 4$ upscaling) associated with the ECCV 2020 workshops. The goal of this challenge is to learn a generic model to super-resolve LR images captured in practical scenarios for more complex degradation than the bicubic down-sampling. In that regard, we propose the SRResCycGAN to super-resolve the LR images with the real-world settings. We use the pretrained model $\mathbf{G_{SR}}$ taken from the SRResCGAN~\cite{muhammad2020srrescgan} (excellent perceptual quality) and further fine-tune it on the training data provided in the AIM 2020 Real Image SR challenge with the proposed SR scheme as shown in the Fig.~\ref{fig:srrescycgan} by using the following training losses:
\begin{equation}
    \mathcal{L}_{G_{SR}} = \mathcal{L}_{\mathrm{GAN}} + \mathcal{L}_{tv} + 10\cdot \mathcal{L}_{\mathrm{1}} + \mathcal{L}_{ssim} + \mathcal{L}_{msssim} +  10\cdot \mathcal{L}_{\mathrm{cyc}}
    \label{eq:l_g_risr}
\end{equation}
Since the final ranking is based on the weighted score of the PSNR and SSIM given in this challenge, we adopt the above losses combination where we neglect the $\mathcal{L}_{\mathrm{per}}$ and use the $\mathcal{L}_{ssim}$ and $\mathcal{L}_{msssim}$ (refers to the Eq.~\eqref{eq:l_g}) whose incorporate the structure similarity~\cite{Wang2004ImageQA} as well as the variations of image resolution and viewing conditions for the output image. Table~\ref{tab:track3} provides the final $\times4$ SR testset results for the track-3 of our method (\textbf{MLP\_SR}) with others participants. We also provide the visual comparison of our method with the state-of-art methods on the track-3 validation-set and testset in the Fig.~\ref{fig:4x_result_val} and Fig.~\ref{fig:4x_result_test}. Our method produces sharp images without any visible corruptions and achieves comparable visual results with the other methods. 

\begin{table}[!ht]
	\centering
	\vspace{-0.5cm}
    \caption{This table reports the quantitative results of our method over the DIV2K validation set (100 images) with unknown degradation for our ablation study. The arrows indicate if high $\uparrow$ or low $\downarrow$ values are desired.}
    \vspace{0.1cm}
	\tabcolsep=0.01\linewidth
	\scriptsize
	\resizebox{1.0\textwidth}{!}{
	\begin{tabular}{|c|c|c|c|c|c|}
		 SR method & Cyclic Path & Network structure  & PSNR$\uparrow$ & SSIM$\uparrow$ & LPIPS$\downarrow$ \\ 
		\hline
		SRResCycGAN & \texttimes & $\by\rightarrow\mathbf{G_{SR}}\rightarrow\hat{\by}$ & 25.05 & 0.67 & \textbf{0.3357}  \\
		SRResCycGAN & \checkmark & $\by\rightarrow\mathbf{G_{SR}}\rightarrow\hat{\by}\rightarrow\mathbf{G_{LR}\rightarrow\by^\prime}$ & 26.13 &  0.71 & 0.3911  \\
		SRResCycGAN+ & \checkmark & $\by\rightarrow\mathbf{G_{SR}}\rightarrow\hat{\by}\rightarrow\mathbf{G_{LR}\rightarrow\by^\prime}$ & \textbf{26.39} &  \textbf{0.73} & 0.4245  \\
	\end{tabular}}
	\label{tab:ablation_study}
	\vspace{-0.5cm}
\end{table}
\subsection{Ablation Study}
For our ablation study, we design two variants of the proposed network structure with cyclic path or not. The first network structure (\ie $\by\rightarrow\mathbf{G_{SR}}\rightarrow\hat{\by}$) takes the LR input to the $\mathbf{G_{SR}}$ and produces the SR output by the supervision of the SR discriminator network $\mathbf{D_{x}}$ without the cyclic path ($\mathbf{G_{LR}}$ \& $\mathbf{D_{y}}$) as shown in the Fig.~\ref{fig:srrescycgan}. Correspondingly, we minimize the total loss in the Eq.~\eqref{eq:l_g} without the $\mathcal{L}_{\mathrm{cyc}}$. The second network structure (\ie $\by\rightarrow\mathbf{G_{SR}}\rightarrow\hat{\by}\rightarrow\mathbf{G_{LR}\rightarrow\by^\prime}$) takes the LR input to the $\mathbf{G_{SR}}$ and produces the SR output by the supervision of the SR discriminator network $\mathbf{D_{x}}$. After that, the SR output fed into the $\mathbf{G_{LR}}$ and reconstructs the LR output by the supervision of the LR discriminator network $\mathbf{D_{y}}$, refers to the Fig.~\ref{fig:srrescycgan}. Accordingly, we minimize the the total loss in the Eq.~\eqref{eq:l_g}.   
Table~\ref{tab:ablation_study} shows the quantitative results of our method over the DIV2K validation-set~\cite{NTIRE2020RWSRchallenge} with the unknown degradation. We found that in the presence of the cyclic path, we get the significant improvement of the PSNR/SSIM \ie $+1.34/+0.06$ to the first variant. It suggests that the cyclic structure gives the benefits to handle complex degradation such as noise, blurring, compression artifacts, etc., while the other structure lacks this due to the domain difference between LR and HR.  

\section{Conclusion}
We proposed a deep SRResCycGAN method for the real image super-resolution problem by handling the domain consistency between the LR and HR images with the CycleGAN. The proposed method solves the SR problem in a GAN framework by minimizing the loss function with the discriminative and residual learning approaches. Our method achieves excellent SR results in terms of the PSNR/SSIM values as well as visual quality compared to the existing state-of-art methods. The SR network is easy to deploy for limited memory storage and CPU power requirements for the mobile/embedded environment. 

%
%
\bibliographystyle{splncs04}
\bibliography{egbib}
\end{document}